\documentclass[reprint, longbibliography, superscriptaddress]{revtex4-1}
\usepackage{graphicx, graphics}
\usepackage{array}
\usepackage{wrapfig}
\usepackage[]{natbib} 
\usepackage{hyperref}
\usepackage{amsmath, amssymb}
\usepackage{
multirow, makecell}
\usepackage{xcolor}

\makeatletter
\renewcommand\frontmatter@abstractwidth{\dimexpr\textwidth-1in\relax}
\makeatother

\begin{document}

\title{Quantifying critical thinking: Development and validation of the Physics Lab Inventory of Critical thinking (PLIC)}
\author{Cole Walsh}
\affiliation{Laboratory of Atomic and Solid State Physics, Cornell University, Ithaca, NY 14853, cjw295@cornell.edu}
\author{Katherine N. Quinn}
\affiliation{Laboratory of Atomic and Solid State Physics, Cornell University, Ithaca, NY 14853, cjw295@cornell.edu}
\author{C. Wieman}
\affiliation{Department of Physics and Graduate School of Education, Stanford University, Stanford, CA 94305}
\author{N.G. Holmes}
\affiliation{Laboratory of Atomic and Solid State Physics, Cornell University, Ithaca, NY 14853, cjw295@cornell.edu}

\date{\today}
\begin{abstract}

Introductory physics lab instruction is undergoing a transformation, with increasing emphasis on developing experimentation and critical thinking skills. These changes present a need for standardized assessment instruments to determine the degree to which students develop these skills through instructional labs. In this article, we present the development and validation of the Physics Lab Inventory of Critical thinking (PLIC). We define critical thinking as the ability to use data and evidence to decide what to trust and what to do. The PLIC is a 10-question, closed-response assessment that probes student critical thinking skills in the context of physics experimentation. Using interviews and data from 5584 students at 29 institutions, we demonstrate, through qualitative and quantitative means, the validity and reliability of the instrument at measuring student critical thinking skills. This establishes a valuable new assessment instrument for instructional labs.

\end{abstract}

\maketitle

\section{Introduction}

More than 400,000 undergraduate students enroll in introductory physics courses at post-secondary institutions in the United States each year~\citep{AIPIntro}. In most introductory physics courses, students spend time learning in lectures, recitations, and instructional laboratories (labs). Labs are notably the most resource-intensive of these course components, as they require specialized equipment, space, facilities, and occupy a significant amount of student and instructional staff time~\citep{Sere2002, White1996}. 

In many institutions, labs are traditionally intended to reinforce and supplement learning of scientific topics and concepts introduced in lecture~\citep{Hofstein1982, Hofstein2004, ALR, Millar2004, Sere2002}. Previous work, however, has called into question the benefit of hands-on lab work to verify concepts seen in lecture~\citep{Sere2002, ALR, Millar2004, Wieman2015AJP, Holmes2017, Wilcox2017}. Traditional labs have also been found to deteriorate students' attitudes and beliefs about experimental physics, while labs that aimed to teach skills were found to improve students' attitudes~\citep{Wilcox2017}. 

Recently, there have been calls to shift the goals of laboratory instruction towards experimentation skills and practices~\citep{PCAST, JTUPP, AAPT14}. This shift in instructional targets provides renewed impetus to develop and evaluate instructional strategies for labs. While much discipline-based education research has evaluated student learning from lectures and tutorials, there is far less research on learning from instructional labs~\citep{Docktor2014, ALR, DBER, Hofstein1982, Hofstein2004, Sere2002}. There exist many open research questions regarding what students are currently learning from lab courses, what students could be learning, and how to measure that learning.

The need for large-scale reform in lab courses goes hand-in-hand with the need for validated and efficient evaluation techniques. Just as previous efforts to reform physics instruction have been unified around shared assessments~\citep{Hake1998, Madsen2013, Madsen2015, Freeman2014}, larger scale change in lab instruction will require common assessment instruments to evaluate that change~\citep{DBER, ALR, Hofstein1982, Hofstein2004}. Currently, few research-based and validated assessment instruments exist for labs. The website PhysPort.org~\cite{physport}, an online resource to support physics instructors with research-based teaching resources, has amassed 92 research-based assessment instruments for physics courses as of this publication. Only four are classified as evaluating lab skills, all of which focus on data analysis and uncertainty~\cite{Day2011, Deardorff2001, Allie1998, Slaughter2012}.

To address this need, we have developed the Physics Lab Inventory of Critical thinking (PLIC), an instrument to assess students' critical thinking skills when conducting experiments in physics. The intent was to provide an efficient, standardized way for instructors at a range of institutions to assess their instructional labs in terms of the degree to which they develop students' critical thinking skills. Critical thinking is defined here as the ways in which one uses data and evidence to make decisions about what to trust and what to do. In a scientific context, this decision-making involves interpreting data, drawing accurate conclusions from data, comparing and evaluating models and data, evaluating methods, and deciding how to proceed in an investigation. These skills and behaviors are commonly used by experimental physicists~\cite{Wieman2015}, but are also relevant to students regardless of their future career paths, academic or otherwise~\cite{JTUPP}. Being able to make sense of data, evaluate whether they are reliable, compare them to models, and decide what to do with them is important whether thinking about introductory physics experiments, interpreting scientific reports in the media, or making public policy decisions. This fact is particularly relevant given that only about 2\% of students taking introductory physics courses at US institutions will graduate with a physics degree~\citep{AIPIntro, AIPDegree}.

In this article, we outline the theoretical arguments for the structure of the PLIC, as well as the evidence of validity and reliability through qualitative and quantitative means.

\vspace{-1.5em}
\section{Assessment Framework}

In what follows, we use arguments from previous literature to motivate the structure and format of the PLIC. We also compare the goals of the PLIC to existing instruments.

\vspace{-1em}
\subsection{Design Concepts}

As discussed above, the goals of the PLIC are to measure critical thinking to fill existing needs and gaps in the existing literature on lab courses~\citep{Holmes2015, Wieman2015, Zwickl2014, Zwickl2015} and reports on undergraduate science, technology, engineering, and math (STEM) instruction~\citep{AAPT14, DBER, Docktor2014}. Probing the \textit{what to trust} component of critical thinking involves testing students' interpretation and evaluation of experimental methods, data, and conclusions. Probing the \textit{what to do} component involves testing what students think should be done next with the information provided. 

Critical thinking is considered content-dependent: ``Thought processes are intertwined with what is being thought about''~\citep[p. 22]{Willingham2008}. This suggests that the assessment questions should be embedded in a domain-specific context, rather than be generic or outside of physics. The content associated with that context, however, should be accessible to students with as minimal physics content knowledge as possible to ensure that the instrument is assessing critical thinking rather than students' content knowledge. We chose to use a single familiar context to optimize the use of students' assessment time.

We chose to have students evaluate someone else's data rather than conduct their own experiment (as in a practical lab exam) for several reasons related to the purposes of standardized assessment. First, collecting data either requires common equipment (which creates logistical constraints and would inevitably reduce its accessibility across institutions) or an interactive simulation, which would create additional design challenges. Namely, it is unclear whether a simulation could sufficiently represent authentic measurement variability and limitations of physical models, which are important for evaluating the \textit{what to trust} aspect of critical thinking. Second, by evaluating another person's work, students are less likely to engage in performance goals~\citep{Adams2007, Dweck1986, Dweck1999}, defined as situations ``in which individuals seek to gain favorable judgments of their competence or avoid negative judgments of their competence"~\citep[p. 1040]{Dweck1986}, potentially limiting the degree to which they would be critical in their thinking. Third, it constrains what they are thinking critically about, allowing us to precisely evaluate a limited and well-defined set of skills and behaviors. By interweaving test questions with descriptions of methods and data, the assessment scaffolds the critical thinking process for the students, again narrowing the focus to target particular skills and behaviors with each question. Fourth, by designing hypothetical results that target specific experimental issues, the assessment can look beyond students' declarative knowledge of laboratory skills to study their enacted critical thinking. 

The final considerations were that the instrument be closed-response and freely available, to meet the goal of providing a practically efficient assessment instrument. The need for a closed-response assessment facilitates automated scoring of student work, making it more likely to be used by instructors. A freely available instrument facilitates its use at a wide range of institutions and for a variety of purposes.

\subsection{Existing instruments}
A number of existing instruments assess skills or concepts related to learning in labs, but none sufficiently meet the design concepts outlined in the previous section (see Table~\ref{tab:Diagnostics}).

The specific skills to be assessed by the PLIC are not covered in any single existing instrument (Table~\ref{tab:Diagnostics}, columns 1-4). Several of these instruments are also not physics-specific (Table~\ref{tab:Diagnostics}, column 6). Given that critical thinking is considered content-dependent~\citep{Willingham2008} and assessment questions should be embedded in a domain-specific context, these instruments do not appropriately assess learning in physics labs. The Physics Measurement Questionnaire (PMQ), Lawson test of scientific reasoning, and Critical Thinking Assessment (CAT) use an open-response format (Table~\ref{tab:Diagnostics}, column 5). This places significant constraints on the use of these assessments in large introductory classes, the target audience of the PLIC. A closed-response version of the PMQ was created~\citep{Lippmann2003}, but limited tests of validity or reliability were conducted on the instrument. The Critical Thinking Assessment Test (CAT) and the California Critical Thinking Skills Test are also not freely available to instructors (Table~\ref{tab:Diagnostics}, column 7). The cost associated with using these instruments places significant logistical burden on dissemination and broad use, failing our goal to support instructors at a wide range of institutions. 

\begin{table*}[htbp]
\caption{Summary of the target skills and assessment structure of existing instruments for evaluating lab instruction.}
\begin{center}
\begin{tabular}{p{3.75cm} >{\centering\arraybackslash}p{2cm} >{\centering\arraybackslash}p{2cm} >{\centering\arraybackslash}p{2cm} >{\centering\arraybackslash}p{2cm} | >{\centering\arraybackslash}p{1.5cm} >{\centering\arraybackslash}p{1.5cm} >{\centering\arraybackslash}p{1.5cm}}
\hline
\hline
& \multicolumn{4}{c}{Skills/Concepts assessed} & \multicolumn{3}{c}{Structure/Features}\\\\
Assessment & Evaluating Data & Evaluating Methods & Evaluating Conclusions & Proposing next steps & Closed-response & Physics-specific & Freely available\\
\hline
Lawson Test of Scientific Reasoning~\citep{Lawson1978} & & X & X & & & & X\\\\
Critical Thinking Assessment Test (CAT)~\citep{Stein2006, Stein2007} & X & & X & & & &\\\\
California Critical Thinking Skills Test~\citep{Facione1990} & & &X  & & X & & \\\\
Test of Scientific Literacy Skills (TOSLS)~\citep{Gormally2012} & X & X & X & & X & & X\\\\
Experimental Design Ability Test (EDAT)~\citep{Sirum2011} & & X & & & X & & X\\\\
Biological Experimental Design Concept Inventory (BEDCI)~\citep{Deane2014} & & X & & & X & & X\\\\
Concise Data Processing Assessment (CDPA)~\citep{Day2011} & X & & & & X & X & X\\\\
Data Handling Diagnostic (DHD)~\citep{DHD} & X & & & & X & X & X\\\\
Physics Measurement Questionnaire (PMQ)~\citep{PMQ} & X & X & & & & X & X\\\\
Measurement Uncertainty Quiz (MUQ)~\citep{MUQ} & X & X & & X & X & X & X\\\\
Physics Lab Inventory of Critical Thinking (PLIC) & X & X & X & X & X & X & X\\\\
\hline
\hline
 \end{tabular}
\end{center}
\label{tab:Diagnostics}
\end{table*}

The Measurement Uncertainty Quiz (MUQ) is the closest of these instruments to meeting the PLIC in assessment goals and structure, but the scope of the concepts covered narrowly focuses on issues of uncertainty. For example, respondents taking the MUQ are asked to propose next steps to specifically reduce the uncertainty. The PLIC allows respondents to propose next steps more generally, such as to consider extending the range of the investigation or testing additional variables.

\subsection{PLIC content and format}\label{PLIC}

The PLIC provides respondents with two case studies of groups completing an experiment to test the relationship between the period of oscillation of a mass hanging from a spring given by: 
\begin{equation}
T=2\pi\sqrt{\frac{m}{k}},
\end{equation}

\noindent where $k$ is the spring constant, $m$ is the mass hanging from the spring, and $T$ is the period of oscillation. The model makes a number of assumptions, such as that the mass of the spring is negligible, though these are not made explicit to the respondents.

The mass on a spring content is commonly seen in high school and introductory physics courses, making it relatively accessible to a wide range of respondents. All required information to describe the physics of the problem, including relevant equations, are given at the beginning of the assessment. Additional content knowledge related to simple harmonic motion is not necessary for answering the questions. This claim was confirmed through interviews with students (described in Secs.~\ref{Construct} and~\ref{Interviews}). This mass on a spring scenario is also rich in critical thinking opportunities, which we had previously observed during real lab situations~\citep{HolmesPhD}.

Respondents are presented with the experimental methods and data from two hypothetical groups:

\begin{enumerate}
    \item The first hypothetical group uses a simple experimental design that involves taking multiple repeated measurements of the period of oscillation of two different masses. The group then calculates the means and standard uncertainties in the mean (standard errors) for the two data sets, from which they calculate the values of the spring constant, $k$, for the two different masses.
    \item The second hypothetical group takes two measurements of the period of oscillation at many different masses (rather than repeated measurements at the same mass). The analysis is done graphically to evaluate the trend of the data and compare it to the predicted model. That is, plotting $T^{2}$ vs $m$ should produce a straight line through the origin. The data, however, show a mismatch between the theoretical prediction and the experimental results. The second group subsequently adds an intercept to the model, which improves the quality of the fit, raising questions about the idealized assumptions in the given model.
\end{enumerate}

\begin{table*}[htbp]
\caption{Question formats used on the PLIC along with the types and examples of questions for each format.
\label{tab:Questions}}
\begin{ruledtabular}
\begin{tabular}{l l l}
\textbf{Format} & \textbf{Types} & \textbf{Examples}\\
\hline
\multirow{2}{*}{Likert} & Evaluate Data & How similar or different do you think Group 1's spring constant (k) values are?\\
& Evaluate Methods & How well do you think Group 1's method tested the model?\\
\hline
\multirow{2}{*}{Multiple choice} & Compare fits & Which fit do you think Group 2 should use?\\
& Compare groups & Which group do you think did a better job of testing the model?\\
\hline
\multirow{2}{*}{Multiple response} & Reasoning & What features were most important in comparing the fit to the data?\\
& What to do next & What do you think Group 2 should do next?
\end{tabular}
\end{ruledtabular}
\end{table*}

PLIC questions are presented on four pages: one page for Group 1, two pages for Group 2 (one with the one-parameter fit, and the other adding the variable intercept to their fit), and one page comparing the two groups. There are a combination of question formats, including five-point Likert-scale questions, traditional multiple choice questions, and \textit{multiple response} questions. Examples of the different question formats are presented in Table~\ref{tab:Questions}. The multiple choice questions each contain three response choices from which respondents can select one. The \textit{multiple response} questions each contain 7-18 response choices and allow respondents to select up to three. The decision to use this format is discussed further in the next section.

\section{Early evaluations of construct validity}\label{Construct}
An open-response version of the PLIC was iteratively developed using interviews with students and written responses from several hundred students, all enrolled in physics courses at universities and community colleges.  These responses were used to evaluate the construct validity of the assessment, defined as how well the assessment measures critical thinking, as defined above.

Six preliminary think-aloud interviews were conducted with students in introductory and upper-division physics courses, with the aim to evaluate the appropriateness of the mass-on-a-spring context. In all interviews, students demonstrated familiarity with the experimental context and successfully progressed through all the questions. The nature of the questions, such that each is related to but independent of the others, allowed students to engage with subsequent questions even if they struggled with earlier ones. All students expressed familiarity with the equipment, physical models, and methods but did not have explicit experience with evaluating the limitations of the model. The interviews also indicated that students were able to progress through the questions and think critically about the data and model without taking their own data. 

The interviews also revealed a number of ways to improve the instrument by refining wording and the hypothetical scenarios. For example, in an early draft, the survey included only the second hypothetical group who fit their data to a line to evaluate the model. Interviewees were asked what they thought it meant to ``evaluate a model.'' Some responded that to evaluate a model one needed to identify where the model breaks down, while others described obtaining values for particular parameters in the model (in this case, the spring constant). The latter is a common goal of experiments in introductory physics labs~\citep{Wieman2015}.
In response to these variable definitions, the first hypothetical group was added to the survey to better capture students' perspectives of what it meant to evaluate a model (i.e., to find a parameter). This revision also offered the opportunity to explore respondents' thinking about comparing pairs of measurements with uncertainty.

The draft open-response instrument was administered to students at multiple institutions described in Table~\ref{tab:Open-Response} in four separate rounds. After each round of open-response analysis, the survey context, and questions, underwent extensive revision. The surveys analyzed in Rounds 1 and 2 provided evidence of the instrument's ability to discriminate between students' critical thinking levels. For example, in cases where the survey was administered as a post-test, only 36\% of the students attributed the disagreement between the data and the predicted model to a limitation of the model. Most students were unable to identify limitations of the model even after instruction. This provided early evidence of the instrument's dynamic range in discriminating between students' critical thinking levels. The different analyses conducted in Rounds 1 and 2 will be clarified in the next section.

\begin{table}[htbp]
\caption{Summary of the number open-response PLIC surveys that were analyzed for early construct validity tests and to develop the closed-response format. The level of the class (first-year [FY] or beyond-first-year [BFY]) and the point in the semester when the survey was administered are indicated.
\label{tab:Open-Response}}
\begin{ruledtabular}
\begin{tabular}{p{2.6cm} c c c c c c}
\textbf{Institution} & \multirow{2}{1cm}{\textbf{Class Level}} & \textbf{Survey} & \multicolumn{4}{c}{\textbf{Round}} \\
& & & 1 & 2 & 3 & 4\\
\hline\\
\multirow{2}{1.5cm}{Stanford University} & \multirow{2}{*}{FY} & Pre & 31 & - & - & -\\
& & Post & - & 30 & - & -\\\\
\multirow{5}{1.5cm}{University of Maine} & \multirow{3}{*}{FY} & Pre & - & - & - & 170\\
& & Mid & - & - & 120 & 93\\
& & Post & - & - & - & 189\\
& \multirow{2}{*}{BFY} & Pre & 25 & - & - & 1\\
& & Post & - & - & - & 3\\\\
\multirow{2}{1.5cm}{Foothill College} & \multirow{2}{*}{FY} & Pre & - & 19 & 40 & -\\
& & Post & - & 19 & 78 & -\\\\
\multirow{2}{2.5cm}{University of British Columbia} & \multirow{2}{*}{FY} & Pre & 10 & - & - & 107\\
& & Post & - & - & -& 38\\\\
\multirow{2}{2.5cm}{University of Connecticut} & \multirow{2}{*}{FY} & Pre & - & - & - & 22\\
& & Post & - & - & - & 3\\\\
\multirow{4}{1.5cm}{Cornell University} & \multirow{2}{*}{FY} & Pre & - & - & - & 89\\
& & Post & - & - & - & 99\\
& \multirow{2}{*}{BFY} & Pre & - & - & - & 35\\
& & Post & - & - & - & 29\\\\
\multirow{2}{2.6cm}{St.Mary's College of Maryland} & \multirow{2}{*}{FY} & Pre & - & - & - & 2\\
& & Post & - & - & - & 3\\
\end{tabular}
\end{ruledtabular}
\end{table}

Students' written responses also revealed significant conceptual difficulties about measurement uncertainty, consistent with other work~\citep{HolmesTPT, Buffler2001, Buffler2009}. The first question on the draft instrument asked students to list the possible sources of uncertainty in the measurements of the period of oscillation. Students provided a wide range of sources of uncertainty, systematic errors (or systematic effects), and measurement mistakes (or literal errors) in response to that prompt~\citep{BFY15}. It was clear that these questions were probing different reasoning than the instrument intended to measure, and so were ultimately removed.

The written responses also exposed issues with the data used by the two groups. For example, the second group originally measured the time for 50 periods for each mass and attached a timing uncertainty of 0.1 seconds for each measured time (therefore 0.02 seconds per period). Several respondents thought this uncertainty was unrealistically small. Others said that no student would ever measure the time for 50 periods. In the revised version, both groups measure the time for five periods.

The survey was also distributed to 78 experts (faculty, research scientists, instructors, and post-docs) for responses and feedback, leading to additional changes and the development of a scoring scheme (Sec.~\ref{subsec:Scoring}). Experts completing the first version of the survey frequently responded in ways that were consistent with how they would expect their students to complete the experiment, which some described in open-response text boxes throughout the survey. We intended for experts to answer the survey by evaluating the hypothetical groups using the same standards that they would apply to themselves or their peers, rather than to students in introductory labs. To make this intention clearer, we changed the survey's language to present respondents with two case studies of hypothetical ``groups of physicists'' rather than ``groups of students.''

\section{Development of closed-response format}\label{CR_format}

To develop and evaluate the closed-response version of the PLIC, we coded the written responses from the open-response instrument, conducted structured interviews, and used responses from experts to generate a scoring scheme. We also developed an automated system to communicate with instructors and collect responses from students, facilitating ease of use and extensive data collection.

\subsection{Generation of the closed response version}
In the first round of analysis, common student responses from the open-response versions were identified through emergent coding of responses collected from the courses listed in Table~\ref{tab:Open-Response} Round 1 ($n = 66$ test responses). The surveys analyzed in Round 1 only include those given before instruction. In the second round, a new set of responses were coded based on this original list of response choices, and any ideas that fell outside of the list were noted. If new ideas came up several times, they were included in the list of response choices for subsequent survey coding. Combined, open response surveys from 134 students across four institutions were used to generate the initial closed-response response choices (Table~\ref{tab:Open-Response} Rounds 1 and 2). The majority of the open responses coded in Round 2 were captured by the response choices created in Round 1, with only one response per student, on average, being coded as \textit{other}. Half of the \textit{other} responses were uninterpretable or unclear (for example, ``collect more data'' with no indication of whether more data meant additional repeated trials, additional mass values, or additional oscillations per measurement). A third round of open-response surveys were coded (Table~\ref{tab:Open-Response} Round 3) to evaluate whether any of the newly generated response choices should be dropped (because they were not generated by enough students overall), whether response choices could be reworded to better reflect student thinking, or if there were any common response choices missing. Researchers regularly checked answers coded as \textit{other} to ensure no common responses were missing, which led the inclusion of several new response choices.

While it was relatively straight forward to categorize students' ideas into closed-response choices, students typically listed multiple ideas in response to each question. A \textit{multiple response} format was employed to address this issue. Response choices are listed in neutral contexts to address the issue of different respondents preferring positive or negative terms (e.g., \textit{the number of masses}, rather than \textit{many masses} and \textit{few masses}).

The closed-response questions, format, and wording were revised iteratively in response to more student interviews and collection of responses from the preliminary closed-response version, which was administered online to the institutions listed in Table~\ref{tab:Open-Response} Round 4. To ensure reliability across student populations, subsets of these students were randomly assigned an open-response survey. Additional coding of these surveys were checked against the existing closed-response choices to again identify additional response choices that could be included and to compare student responses between open- and closed-response surveys. The number of open-response surveys analyzed from these institutions is summarized in Table~\ref{tab:Open-Response} Round 4.

\subsection{Interview analysis}\label{Interviews}
Two sets of interviews were conducted and audio-recorded to evaluate the closed-response version of the PLIC. A demographic breakdown of students interviewed is presented in Table~\ref{tab:Interviews}.

\begin{table}[htbp]
\caption{Demographic breakdown of students interviewed to probe the validity of the closed-response PLIC.
\label{tab:Interviews}}
\begin{ruledtabular}
\begin{tabular}{c l c c}
 \textbf{Category} & \textbf{Breakdown} & \textbf{Set 1} & \textbf{Set 2}\\
 \hline
Major & Physics or Related Field & 8 & 3 \\
      & Other STEM & 4 & 6 \\
\hline
Academic & Freshman & 6 & 3\\ 
Level    & Sophomore & 3 & 2\\
         & Junior & 1 & 3\\ 
         & Senior & 0 & 1 \\
         & Graduate Student & 2 & 0\\
\hline
Gender & Women & 6 & 5\\ 
       & Men & 6 & 4\\
\hline
Race/          & African-American & 2 & 1 \\
Ethnicity      & Asian/Asian-American & 3 & 3 \\ 
               & White/Caucasian & 5 & 4 \\ 
               & Other & 2 & 1
\end{tabular}
\end{ruledtabular}
\end{table}

In the first set of interviews, participants completed the closed-response version of the PLIC in a think-aloud format. One goal of the interviews was to ensure that the instrument was measuring critical thinking without testing physics content knowledge, using a broader pool of participants. All participants were able to complete the assessment, including non-physics majors, and there was no indication that physics content knowledge limited or enhanced their performance. The interviews identified some instances where wording was unclear, such as statements about residual plots.

A secondary goal was to identify the types of reasoning participants employed while completing the assessment (whether critical thinking or other). We observed three distinct reasoning patterns that participants adopted while completing the PLIC~\cite{Quinn2018}: (1) \textit{selecting all} (or almost all) possible choices presented to them, (2) \textit{cueing} to keywords, and (3) carefully considering and \textit{discerning} the response choices. The third pattern presented the strongest evidence of critical thinking. To reduce the \textit{select all} behavior, respondents are now limited to selecting no more than three response choices per question.

In the second set of interviews, participants completed the open-response version of the PLIC in a think-aloud format, and then were given the closed-response version. The primary goal was to assess the degree to which the closed-response options cued students' thinking. For each of the four pages of the PLIC, participants were given the descriptive information and \textit{Likert} and \textit{multiple choice} questions on the computer, and were asked to explain their reasoning out loud, as in the open-response version. Participants were then given the closed-response version of that page before moving on to the next page in the same fashion. 

When completing the closed-response version, all participants selected the answer(s) that they had previously generated. Eight of the nine participants also carefully read through the other choices presented to them and selected additional responses they had not initially generated. Participants typically generated one response on their own and then selected one or two additional choices. This behavior was representative of the \textit{discerning} behavior observed in the first round of interviews and provided evidence that limiting respondents to selecting no more than three response choices prompts respondents to be discerning in their choices. No participant generated a response that was not in the closed-response version, or expressed a desire to select a response that was not presented to them.

While taking the open-response version, two of the nine participants were hesitant to generate answers aloud to the question ``What do you think Group 2 should do next?,'' however no participant hesitated to answer the closed-response version of the questions. Another participant expressed how hard they thought it was to select no more than three response choices in the closed-response version because ``these are all good choices,'' and so needed to carefully choose response choices to prioritize. It is possible that the combination of open-response questions followed by their closed-response partners could prompt students to engage in more critical thinking than either set of questions alone. Future work will evaluate this possibility. The time associated with the paired set of questions, however, would likely make the assessment too long to meet our goal for an efficient assessment instrument.

\subsection{Development of a scoring scheme}\label{subsec:Scoring}

The scoring scheme was developed through evaluations of 78 responses to the PLIC from expert physicists (faculty, research scientists, instructors, and post-docs). As discussed in Sec.~\ref{PLIC}, the PLIC uses a combination of Likert questions, traditional multiple choice questions, and \textit{multiple response} questions. This complex format meant that developing a scoring scheme for the PLIC was non-trivial.

\subsubsection{Likert and multiple choice questions}

The Likert and multiple choice questions are not scored. Instead, the assessment focuses on students' reasoning behind their answers to the Likert and multiple choice questions. For example, whether a student thinks Group 1's method effectively evaluated the model is less indicative of critical thinking than why they think it was effective.

The Likert and multiple choice questions are included in the PLIC to guide respondents in their thinking on the \textit{reasoning} and \textit{what to do next} questions. Respondents must answer (either implicitly or explicitly) the question of how well a group tested the model before providing their reasoning. The PLIC makes these decision processes explicit through the Likert and multiple choice questions.

\subsubsection{Multiple response}\label{Sec:MultipleResponse}

There were several criteria for determining the scoring scheme for the PLIC. First and foremost, the scoring scheme should align with how experts answer. Expert responses indicated that there was no single correct answer for any of the questions. As indicated in the previous sections, all response choices were generated by students in the open-response versions, and so many of the responses are reasonable choices. For example, when seeing data that is in tension with a given model, it is reasonable to check the assumptions of the model, test other possible variables, or collect more data. An all-or-nothing scoring scheme, where respondents receive full credit for selecting \textit{all} of the correct responses and \textit{none} of the incorrect responses~\citep{butler2018multiple}, was therefore inappropriate. There needed to be multiple ways to obtain full points for each question, in line with how experts responded.

We assign values to each response choice equal to the fraction of experts who selected the response (rounded to the nearest tenth). As an example, the first \textit{reasoning} question on the PLIC asks respondents to identify what features were important for comparing the spring constants, $k$, from the two masses tested by Group 1. About 97\% of experts identified ``the difference between the $k$-values compared to the uncertainty'' (R1) as being important. Therefore, we assign a value of 1 for this response choice. About 32\% identified ``the size of the uncertainty'' (R2) as being important, and so we assign a value of 0.3 for this response choice. All other response choices received support from less than 12\% of experts and so are assigned values of 0 or 0.1, as appropriate. These values are valid in so far as the fraction of experts selecting a particular response choice can be interpreted as the relative correctness of the response choice. These values will continue to be evaluated as additional expert responses are collected.

Another criteria was to account for the fact that respondents may choose as many as zero to three response choices per question. Accordingly, we sum the total value of responses selected and divide by the maximum value of the number of responses selected:

\begin{equation}\label{eq:Scoring}
    Score = \frac{\sum_{n = 1}^{i} V_{n}}{V_{max_{i}}},
\end{equation}
where $V_{n}$ is the value of the $n^{th}$ response choice selected and $V_{max_{i}}$ is the maximum attainable score when $i$ response choices are selected. Explicitly, the values of $V_{max_{i}}$ are:

\begin{equation}
    \begin{split}
        V_{max_{1}} &= \text{Highest Value},\\
        V_{max_{2}} &= (\text{Highest Value}) + (\text{Second Highest Value}),\\
        V_{max_{3}} &= (\text{Highest Value}) + (\text{Second Highest Value})\\
        &+ (\text{Third Highest Value}).
    \end{split}
\end{equation}

If a respondent selects $N$ responses, then they will obtain full credit if they select the $N$ highest valued responses. In the example above, a respondent selecting one response must select R1 in order to receive full credit. A respondent selecting two responses must select both R1 and R2 to receive full credit. A respondent selecting three responses must select R1 and R2, as well as the third highest valued response to receive full credit. 

The scoring scheme rewards picking highly-valued responses more than it penalizes for picking low-valued responses. For example, respondents selecting R1, R2, and a third response choice with value zero will receive a score of 0.93 on this question. The scoring scheme does not necessarily penalize for selecting more or fewer responses. For example, a respondent who picks the three highest-valued responses will receive the same score as a respondent who picks only the two highest-valued responses. This is true even if the third highest-valued response may be considered a poor response (with few experts selecting it). Fortunately, most questions on the PLIC have a viable third response choice.

The respondent's overall score on the PLIC is obtained by summing scores on each of the \textit{multiple response} questions. Thus, the maximum attainable score on the PLIC is 10 points. Using this scheme, the 78 experts obtained an average overall score of $7.6\pm0.2$. The distribution of these scores are shown in Fig.~\ref{Maturity_Box} in comparison to student scores (discussed in Sec.~\ref{Concurrent_Validity}). Here and throughout, uncertainties in our results are given by the 95\% confidence interval (i.e. 1.96 multiplied by the standard error of the quantity).

\subsection{Automated Administration}

The PLIC is administered online via Qualtrics as part of an automated system adapted from Ref.~\citep{wilcox2016a}. Instructors complete a Course Information Survey (CIS) through a web link (available at~\citep{cperl}) and are subsequently sent a unique link to the PLIC for their course. The system handles reminder emails, updates, and the activation and deactivation of pre- and post-surveys using information provided by the instructor in the CIS. Instructors are also able to update the activation and deactivation dates of one or both of the pre- and post-surveys via another link hosted on the same webpage as the CIS. Following the deactivation of the post-survey link, instructors are sent a summary report detailing the performance of their class compared to classes of a similar level.

\section{Statistical reliability testing}\label{sec:Stats}

Below, we use classical test theory to investigate the reliability of the assessment, including test and question difficulty, time to completion reliability, the discrimination of the instrument and individual questions, the internal consistency of the instrument, test-retest reliability, and concurrent validity.

\subsection{Data Sources}

We conducted statistical reliability tests on data collected using the most recent version of the instrument. These data include students from 27 different institutions and 58 distinct courses who have taken the PLIC \textit{at least once} between August 2017 and December 2018. We had 41 courses from PhD granting institutions, 4 from master's granting institutions, and 13 from two or four-year colleges. The majority of the courses (33) were at the first-year level with 25 at the beyond-first-year level.

Only valid student responses were included in the analysis. To be considered valid, a respondent must:

\begin{enumerate} \setlength\itemsep{0em}
    \item click submit at the end of the survey,
    \item consent to participate in the study,
    \item indicate that they are at least 18 years of age,
    \item and spend at least 30 seconds on at least one of the four pages.
\end{enumerate}

The time cut-off (criteria 4) was imposed because it took a typical reader approximately 20 seconds to randomly click through each page, without reading the material. Students who spent too little time could not have made a legitimate effort to answer the questions. Of these valid responses, pre- and post-responses were matched for individual students using the student ID and/or full name provided at the end of the survey. The time cut-off removed 181 (7.6\%) students from the matched dataset.

We collected \textit{at least one} valid survey from 4329 students and matched pre- and post-surveys from 2189 students. The demographic distribution of these data are shown in Table~\ref{tab:Breakdown}. We have collapsed all physics, astronomy, and engineering physics majors into the ``physics'' major category. Instructors reported the estimated number of students enrolled in their classes, which allow us to estimate response rates. The mean response rate to the pre-survey was $64\pm4\%$, while the mean response rate to the post-survey was $49\pm4\%$.

\begin{table}[b]
\caption{Percentage of respondents in valid pre-, post-, and matched datasets broken down by gender, ethnicity, and major. Students had the option to not disclose this information, so percentages may not sum to 100\%.\label{tab:Breakdown}}
\begin{ruledtabular}
\begin{tabular}{l c c c}
& \textbf{Pre} & \textbf{Post} & \textbf{Matched}\\
\textbf{Total} & 3635 & 2883 & 2189\\
\hline
\textbf{Gender} & & &\\
Women & 39\% & 39\% & 40\%\\
Men & 59\% & 60\% & 59\%\\
Other & 0.5\% & 0.3\% & 0.4\%\\
\textbf{Major} & & &\\
Physics & 18\% & 19\% & 20\%\\
Engineering & 43\% & 44\% & 45\%\\
Other Science & 31\% & 29\% & 29\%\\
Other & 5.7\% & 5.8\% & 5.7\%\\
\textbf{Race/Ethnicity} & & &\\
American Indian & 0.7\% & 0.6\% & 0.5\%\\
Asian & 25\% & 26\% & 28\%\\
African American & 2.8\% & 2.6\% & 2.4\%\\
Hispanic & 4.3\% & 5.3\% & 4.8\%\\
Native Hawaiian & 0.4\% & 0.4\% & 0.4\%\\
White & 62\% & 61\% & 61\%\\
Other & 1.3\% & 1.6\% & 1.2\%
\end{tabular}
\end{ruledtabular}
\end{table}

As part of the validation process, 20\% of respondents were randomly assigned open-response versions of the PLIC during the 2017-2018 academic year. As such, some students in our matched dataset saw both a closed-response and open-response version of the PLIC. Table~\ref{tab:matched_dataset} presents the number of students in the matched dataset that saw each version of the PLIC at pre- and post-instruction. Additional analysis of the open-response data will be included in a future publication, but in the analyses presented here, we focus solely on the closed-response version of the PLIC (1911 students completed both a closed-response pre-survey and a closed-response post-survey).

\begin{table}[b]
\caption{Number of students in the matched dataset who took each version of the PLIC. The statistical analyses focus exclusively on students who completed both a closed-response pre-survey and closed-response post-survey.\label{tab:matched_dataset}}
\begin{ruledtabular}
\begin{tabular}{l c c}
\textbf{Pre-survey version} & \textbf{Post-survey version} & \textbf{N}\\
\hline
Closed-response & Closed-response & 1911\\
 & Open-response & 105\\
Open-response & Closed-response & 144\\
 & Open-response & 29\\
\end{tabular}
\end{ruledtabular}
\end{table}

\subsection{Test and question scores}\label{Test_Scores}

\begin{figure}
\includegraphics[width = \linewidth]{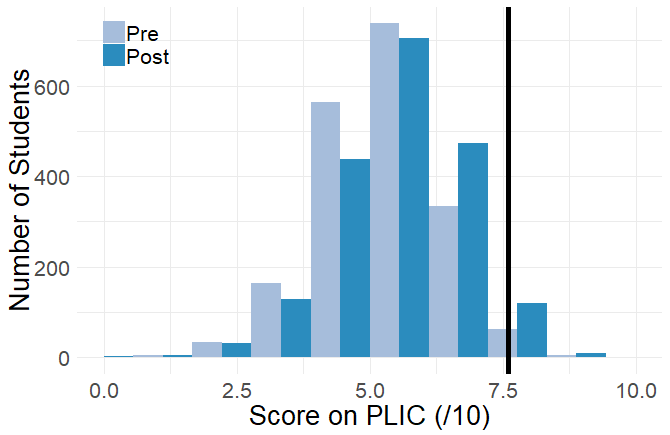}
\caption{Distributions of respondents' matched pre- and post-scores on the PLIC ($N = 1911$). The average score obtained by experts is marked with a black vertical line.
\label{fig:Distribution}}
\end{figure}

In Fig.~\ref{fig:Distribution} we show the matched pre- and post-survey distributions ($N = 1911$) of respondents' total scores on the PLIC. The average total score on the pre-survey is $5.25\pm0.05$ and the average score on the post-survey is $5.52\pm0.05$. The data follow an approximately normal distribution with roughly equal variances in pre- and post-scores. For this reason, we use parametric statistical tests to compare paired and unpaired sample means. The pre- and post-survey means are statistically different (paired $t$-test, $p < 0.001$) with a small effect size (Cohen's $d = 0.23$).

\begin{figure}
\includegraphics[width = \linewidth]{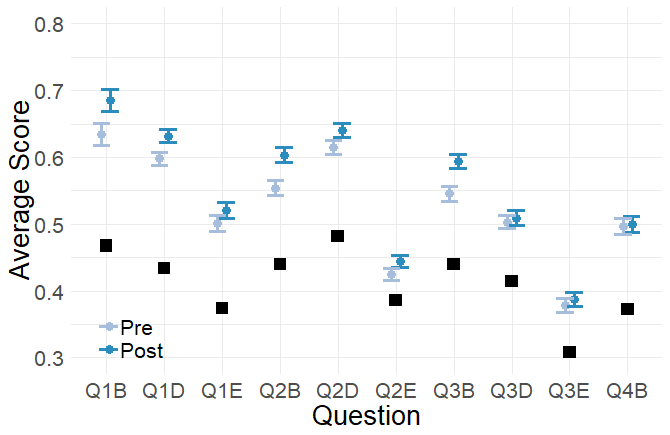}
\caption{Average score (representing question difficulty) for each of the 10 PLIC questions. The average expected score that would be obtained through random guessing is marked with a black square. Scores on the first seven questions are statistically different between the pre- and post-surveys (Mann-Whitney U, Holm-Bonferroni corrected $\alpha < 0.05$).
\label{fig:Question_Difficulty}}
\end{figure}

In addition to overall PLIC scores, we also examined scores on individual questions (Fig.~\ref{fig:Question_Difficulty}, i.e., question difficulty). All questions differ in the number of closed-response choices available and the score values of each response. We simulated 10,000 students randomly selecting three responses for each question to determine a baseline difficulty for each question. These random guess scores are indicated as black squares in Fig.~\ref{fig:Question_Difficulty}.

The average score per question ranges from 0.30 to 0.80, within the acceptable range for educational assessments~\citep{Ding2009, Doran1980}. Correcting \textit{p}-values for multiple comparisons using the Holm-Bonferroni method, each of the first seven questions showed statistically significant increases between pre- and post-test at the $\alpha = 0.05$ significance level.

The two questions with the lowest average score (Q2E and Q3E) correspond to \textit{what to do next} questions for Group 2. These questions correspond to two of the four lowest scores through random guessing. Furthermore, the highest valued responses on these questions involve changing the fit line, investigating the non-zero intercept, testing other variables, and checking the assumptions of the model. It is not surprising that students have low scores on these questions, as they are seldom exposed to this kind of investigation, particularly in traditional labs~\citep{Holmes2015}. Although these average scores are low (question difficulty is high), they are still within the acceptable range of [0.3, 0.8] for educational assessments~\citep{Ding2009, Doran1980}. 

\subsection{Time to completion}\label{Time}

We examined the relationship between a respondent's score and the amount of time they took to complete the PLIC. The total assessment duration was defined as the time elapsed from the moment a respondent opened the survey to the moment they submitted it. This time then encompasses any time that the respondent spent away or disengaged from the survey.

The median time for completing the PLIC is $16.0\pm 0.4$ minutes for the pre-survey and $11.5\pm 0.3$ minutes for the post-survey. The correlation between total PLIC score and time to completion is $r < 0.03$ for both the pre-survey and the post-survey, suggesting that there is no relationship between a student's score on the PLIC and the time they take to complete the assessment.

\subsection{Discrimination}\label{Discrimination}

Ferguson's $\delta$ coefficient gives an indication of how well an instrument discriminates between individual students by examining how many unequal pairs of scores exist. When scores on an assessment are uniformly distributed, this index is exactly 1~\citep{Ding2009}. A Ferguson's $\delta$ greater than 0.90 indicates good discrimination among students. Because the PLIC is scored on a nearly continuous scale, all but three students received a unique score on the pre-survey, and all but two students received a unique score on the post-survey. Ferguson's $\delta$ is equal to 1 for both the pre- and post-surveys.

We also examined how well each question discriminates between high and low performing students using question-test correlations (that is, correlations between respondents' scores on individual questions and their score on the full test). Question-test correlations are greater than 0.38 for all questions on the pre-survey and greater than 0.42 for all questions on the post-survey. None of these question-test correlations are statistically different between the pre- and post-surveys (Fisher transformation) after correcting for multiple testing effects (Holm-Bonferroni $\alpha < 0.05$). All of these correlations are well above the generally accepted threshold for question-test correlations, $r \geq 0.20$~\citep{Ding2009}.

\subsection{Internal Consistency}\label{Internal_Consistency}

While the PLIC was designed to measure critical thinking, our definition of critical thinking demonstrates that it is not a unidimensional construct. To confirm that the PLIC does indeed exhibit this underlying structure, we performed a Principal Component Analysis (PCA) on question scores. PCA is a dimensionality reduction technique that finds groups of question scores that vary together. In a unidimensional assessment, all questions should group together, and so there would only be one principal component to explain most of the variance.

We performed a PCA on matched pre-surveys and post-surveys and found that six principal components were needed to explain at least 70\% of the variance in 10 PLIC question scores in both cases. The first three principal components can explain 45\% of the variance in respondents' scores. It is clear, therefore, that the PLIC is not a single construct assessment. The first three principal components are essentially identical between the pre- and post-surveys (their inner products between the pre- and post-survey are 0.98, 0.97, and 0.94, respectively).

Cronbach's $\alpha$ is typically reported for diagnostic assessments as a measure of the internal consistency of the assessment. However, as argued elsewhere~\citep{Day2011, Wilcox2016}, Cronbach's $\alpha$ is primarily designed for single construct assessments and depends on the number of questions as well as the correlations between individual questions.

We find that \smash{$\alpha = 0.54$} for the pre-survey and $\alpha=0.60$ for the post-survey, well below the suggested minimum value of $\alpha = 0.80$~\citep{Engelhardt2009} for a single construct assessment. We conclude, then, in accordance with the PCA results above, that the PLIC is not measuring a single construct and Cronbach's $\alpha$ cannot be interpreted as a measure of internal reliability of the assessment.

\subsection{Test-retest reliability}\label{Test-retest}

The repeatability or test-retest reliability of an assessment concerns the agreement of results when the assessment is carried out under the same conditions. The test-retest reliability of an assessment is usually measured by administering the assessment under the same conditions multiple times to the same respondents. Because students taking the PLIC a second time had always received some physics lab instruction, it was not possible to establish test-retest reliability by examining the same students' individual scores. Instead, we estimated the test-retest reliability of the PLIC by comparing the pre-survey scores of students in the same course in different semesters. Assuming that students attending a particular course in one semester are from the same population as those who attend that same course in a later semester, we expect pre-instruction scores for that course to be the same in both semesters. This serves as a measure of the test-retest reliability of the PLIC at the course level rather than the student level.

In all, there were six courses from three institutions where the PLIC was administered prior to instruction in at least two separate semesters, including two courses where it was administered in three separate semesters. We performed ANOVA evaluating the effect of semester on average pre-score and report effect sizes as \textit{$\eta^{2}$} for each of the six courses. When the between-groups degrees of freedom are equal to one (i.e., there are only two semesters being compared), the results of the ANOVA are equal to those obtained from an unpaired $t$-test with $F = t^{2}$. The results are shown in Table~\ref{tab:Test-Retest}.

\begin{table}[b]
\caption{Summary of test-retest results comparing pre-surveys from multiple semesters of the same course.
\label{tab:Test-Retest}}
\begin{ruledtabular}
\begin{tabular}{l c c c c}
\textbf{Class} & \textbf{Semester} & \textbf{N} & \textbf{Pre Avg.} & \textbf{Comparisons}\\ 
\hline
& Fall 2017 & 59 & $5.2\pm0.3$ & $F(2, 363) = 0.192$\\
Class A & Spring 2018 & 92 & $5.3\pm0.2$ & $p = 0.826$\\
& Fall 2018 & 215 & $5.24\pm0.14$ & $\eta^{2} = 0.001$\\
\hline
& Fall 2017 & 79 & $5.5\pm0.2$ & $F(2, 203) = 0.417$\\
Class B & Spring 2018 & 36 & $5.7\pm0.4$ & $p = 0.660$\\
& Fall 2018 & 91 & $5.5\pm0.2$ & $\eta^{2} = 0.004$\\
\hline
\multirow{3}{*}{Class C} & \multirowcell{3}{Fall 2017\\Fall 2018} & \multirowcell{3}{90\\119} & \multirowcell{3}{$5.8\pm0.3$\\$5.50\pm0.17$} & $F(1, 207) = 4.54$ \\
& & & & $p = 0.034$\\
& & & & $\eta^{2} = 0.021$\\
\hline
\multirow{3}{*}{Class D} & \multirowcell{3}{Spring 2018\\Fall 2018} & \multirowcell{3}{40\\16} & \multirowcell{3}{$6.3\pm0.3$\\$6.2\pm0.7$} & $F(1, 54) = 0.054$\\
& & & & $p = 0.818$\\
& & & & $\eta^{2} = 0.001$\\
\hline
\multirow{3}{*}{Class E} & \multirowcell{3}{Fall 2017\\Fall 2018} & \multirowcell{3}{142\\289} & \multirowcell{3}{$5.0\pm0.2$\\$4.79\pm0.12$} & $F(1, 429) = 1.37$\\
& & & & $p = 0.153$\\
& & & & $\eta^{2} = 0.005$\\
\hline
\multirow{3}{*}{Class F} & \multirowcell{3}{Spring 2018\\Fall 2018} & \multirowcell{3}{95\\89} & \multirowcell{3}{$5.0\pm0.2$\\$5.1\pm0.2$} & $F(1, 182) = 1.89$\\
& & & & $p = 0.171$\\
& & & & $\eta^{2} = 0.010$\\
\end{tabular}
\end{ruledtabular}
\end{table}

The p-values indicate that pre-survey means were not statistically different between semesters for any class other than Class C, but these p-values have not been corrected for multiple comparisions. After correcting for multiple testing effects using either the Bonferroni or Holm-Bonferroni method at $\alpha < 0.05$ significance, pre-survey means were not statistically different for Class C.

Given the possibility of small variations in class populations from one semester to the next, it is reasonable to expect small effects to arise on occasion, so the moderate difference in pre-survey means for Class C is not surprising. 
As we collect data from more classes, we will continue to check the test-retest reliability of the instrument in this manner. 

\subsection{Concurrent validity}\label{Concurrent_Validity}
We define concurrent validity as a measure of the consistency of performance with expected results. For example, we expect that either from instruction or selection effects, performance on the PLIC should increase with greater physics maturity of the respondent. We define physics maturity by the level of the lab course that respondents were enrolled in when they took the PLIC.

\begin{figure}
\includegraphics[width = \linewidth]{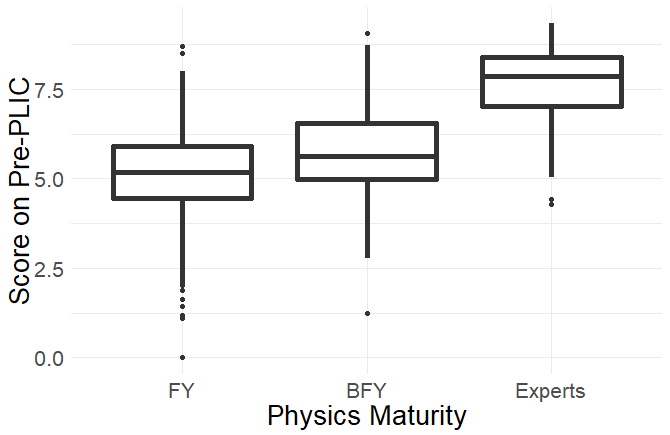}
\caption{Boxplots comparing pre-survey scores for students in first-year (FY) and beyond-first-year (BFY) labs, and experts. Horizontal lines indicate the lower and upper quartiles and the median score. Scores lying outside $1.5\times IQR$ (Inter-Quartile Range) are indicated as outliers.
\label{Maturity_Box}}
\end{figure}

To assess this form of concurrent validity, we split our matched dataset by physics maturity. This split dataset included 1558 respondents from first-year (FY) labs, 353 respondents from beyond-first-year (BFY) labs, and 78 physics experts. Figure ~\ref{Maturity_Box} compares the performance on the pre-survey by physics maturity of the respondent. In Table~\ref{Score_Level}, we report the average scores for respondents for both the pre- and post-surveys across maturity level. The significance level and effect sizes between pre- and post-mean scores within each group are also indicated.

\begin{table}[htbp]
\caption{Average scores across levels of physics maturity, where \textit{N} is the number of matched responses, except for experts who only filled out the survey once. Significance levels and effect sizes are reported for differences between the pre- and post-survey means. (FY = first-year lab course, BFY = beyond-first-year lab course) \label{Score_Level}}
\begin{ruledtabular}
\begin{tabular}{cl c c c c}
& \textit{\textbf{N}} & \textbf{Pre Avg.} & \textbf{Post Avg.} & \textit{\textbf{p}} & \textit{\textbf{d}}\\ 
\hline
FY & 1558 & $5.15\pm0.05$ & $5.45\pm0.06$ & $<0.001$ & 0.215\\
BFY & 353 & $5.69\pm0.12$ & $5.81\pm0.13$ & 0.095 & 0.089\\
Experts & 78 & $7.6\pm0.2$ & & \\
\end{tabular}
\end{ruledtabular}
\end{table}

There is a statistically significant difference between pre- and post-survey means for students trained in FY labs with a small effect size, but not in BFY labs.

An ANOVA comparing pre-survey scores across physics maturity (FY, BFY, expert) indicates that physics maturity is a statistically significant predictor of pre-survey scores ($F(2, 1986) = 203$, $p < 0.001$)  with a large effect size ($\eta^{2} = 0.170$). The large differences in means between groups of differing physics maturity, coupled with the small increase in mean scores following instruction at both the FY and BFY level, may imply that these differences arise from selection effects rather than cumulative instruction. This selection effect has been seen in other evaluations of students' lab sophistication as well~\citep{Wilcox2017b}.

\begin{figure}
\includegraphics[width = \linewidth]{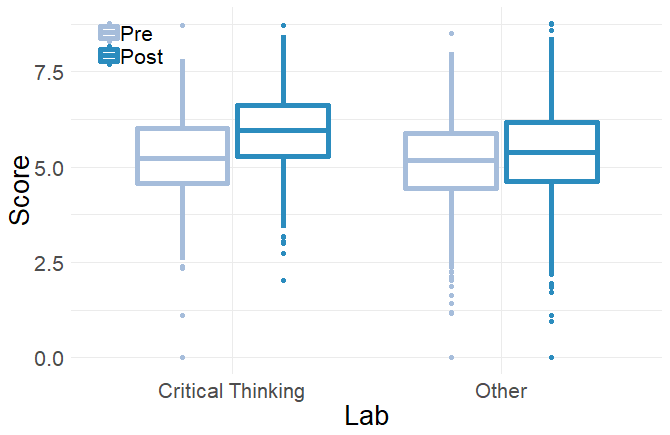}
\caption{Boxplots of students' total scores on the PLIC grouped by the type of lab students participated in. Horizontal lines indicate the lower and upper quartiles and the median score. Scores lying outside $1.5\times IQR$ (Inter-Quartile Range) are labelled as outliers.
\label{Lab_Box}}
\end{figure}

Another measure of concurrent validity is through the impact of lab courses that aim to teach critical thinking on student performance. We grouped FY students according to the type of lab their instructor indicated they were running as part of the Course Information Survey. The data include 273 respondents who participated in FY labs designed to teach critical thinking as defined for the PLIC (CTLabs) and 1285 respondents who participated in other FY physics labs. Boxplots of these scores split by lab type are shown in Fig.~\ref{Lab_Box}.

We fit a linear model predicting post-scores as a function of lab treatment and pre-score:

\begin{equation}\label{eq:LM}
    PostScore_{i} = \beta_{0} + \beta_{1}*CTLabs_{i} + \beta_{2}*PreScore_{i}.
\end{equation}
The results are shown in Table~\ref{tab:Lab}. We see that, controlling for pre-scores, lab treatment has a statistically significant impact on post-scores; students trained in CTLabs perform 0.52 points higher, on average, at post-test compared to their counterparts trained in other labs with the same pre-score.

\begin{table}[htbp]
\caption{Linear model for PLIC post-score as a function of lab treatment and pre-score.\label{tab:Lab}}
\begin{ruledtabular}
\begin{tabular}{l c c c}
\textbf{Variable} & \textbf{Coefficient} & \textit{\textbf{p}}\\ 
\hline
Constant, $\beta_{0}$ & $4.6\pm0.3$ & $<0.001$\\
CTLabs, $\beta_{1}$ & $0.52\pm0.15$ & $<0.001$\\
Pre-score, $\beta_{2}$ & $0.25\pm0.05$ & $<0.001$\\
\end{tabular}
\end{ruledtabular}
\end{table}

\subsection{Partial sample reliability}\label{Partial_Sample}
The partial sample reliability of an assessment is a measure of possible systematics associated with selection effects (given that response rates are less than 100\% for most courses)~\citep{Wilcox2016}. We compared the performance of matched respondents (those who completed a closed-response version of both the pre- and post-survey) to respondents who completed only one closed-response survey. The results are summarized in Table~\ref{tab:Partial}. Means are statistically different between the matched and unmatched datasets for both the pre- and post-survey.

\begin{table}[htbp]
\caption{Sscores from respondents who took both a closed-response pre- and post-survey (matched dataset) and students who only took one closed-response survey (unmatched dataset). $p$-values and effect sizes $d$ are reported for the differences between the matched and unmatched datasets in each case.\label{tab:Partial}}
\begin{ruledtabular}
\begin{tabular}{l c c c c c c}
\textbf{Survey} & \textbf{Dataset} & \textbf{N} & \textbf{Avg.} & \textit{\textbf{p}} & \textit{\textbf{d}}\\ 
\hline
\multirow{2}{*}{Pre} & Matched & 1911 & $5.25\pm0.05$ & \multirow{2}{*}{0.011} & \multirow{2}{*}{0.090}\\
& Unmatched & 1376 & $5.15\pm0.06$ & &\\
\multirow{2}{*}{Post} & Matched & 1911 & $5.52\pm0.05$ & \multirow{2}{*}{$<0.001$} & \multirow{2}{*}{0.242}\\
& Unmatched & 797 & $5.22\pm0.09$ & &\\
\end{tabular}
\end{ruledtabular}
\end{table}

These biases, though small, may still have a meaningful impact on our analyses, given that we have neglected certain lower performing students (particularly for the tests of concurrent validity). We address these potential biases by imputing our missing data in Appendix~\ref{appendix}. This analysis indicates that the missing students did not change the results.

\section{Summary and conclusions}

We have presented the development of the PLIC for measuring student critical thinking, including tests of its validity and reliability. We have used qualitative and quantitative tools to demonstrate the PLIC's ability to measure critical thinking (defined as the ways in which one makes decisions about what to trust and what to do) across instruction and general physics training. There are several implications for researchers and instructors that follow from this work. 

For researchers, the PLIC provides a valuable new tool for the study of educational innovations. While much research has evaluated student conceptual gains, the PLIC is an efficient and standardized way to measure students' critical thinking in the context of introductory physics lab experiments. As the first such research-based instrument in physics, this facilitates the exploration of a number of research questions. For example, do the gender gaps common on concept inventories~\citep{Madsen2013} exist on an instrument such as the PLIC? How do different forms of instruction impact learning of critical thinking? How do students' critical thinking skills correlate with other measures, such as grades, conceptual understanding, persistence in the field, or attitudes towards science?

For instructors, the PLIC now provides a unique means of assessing student success in introductory and upper-division labs. As the goals of lab instruction shift from reinforcing concepts to developing experimentation and critical thinking skills, the PLIC can serve as a much needed research-based instrument for instructors to assess the impact of their instruction. Based on the discrimination of the instrument, its use is not limited to introductory courses. The automated delivery system is straightforward for instructors to use. The vast amounts of data already collected also allow instructors to compare their classes to those across the country. 

In future studies related to PLIC development and use, we plan to evaluate patterns in students' responses through cluster and network analysis. Given the use of multiple response questions and similar questions for the two hypothetical groups, these techniques will be useful in identifying related response choices within and across questions. Also, the analyses thus far have largely ignored data collected from the open-response version of the PLIC. Unlike the closed-response version, the open-response version has undergone very little revision over the course of the PLIC's development and we have collected over 1000 responses. We plan to compare responses to the open- and closed-response versions to evaluate the different forms of measurement. The closed-response version measures students' abilities to critique and choose among multiple ideas (accessible knowledge), while the open-response version measures students' abilities to generate these ideas for themselves (available knowledge). The relationship between the accessibility and availability of knowledge has been studied in other contexts~\citep{heckler2018reasoning} and it will be interesting to explore this relationship further with the PLIC.

\section{Acknowledgements}
This material is based upon work supported by the National Science Foundation under Grant No. 1611482. We are grateful to undergraduate researchers Saaj Chattopadhyay, Tim Rehm, Isabella Rios, Adam Stanford-Moore, and Ruqayya Toorawa who assisted in coding the open-response surveys. Peter Lepage, Michelle Smith, Doug Bonn, Bethany Wilcox, Jayson Nissen, and members of CPERL provided ideas and useful feedback on the manuscript and versions of the instrument. We are very grateful to Heather Lewandowski and Bethany Wilcox for their support developing the administration system. We would also like to thank all of the instructors who have used the PLIC with their courses, especially Mac Stetzer, David Marasco, and Mark Kasevich for running the earliest drafts of the PLIC. 

\appendix
\section{Multiple Imputation Analysis}\label{appendix}
In the analyses in the main text, we used only PLIC data from respondents who had taken the closed-response PLIC at both pre- and post-survey. As discussed in Sec.~\ref{Partial_Sample}, the missing data biases the average scores toward higher scores. It is likely that this bias is due to a skew in representation toward students who receive higher grades in the matched dataset compared to the complete dataset~\citep{nissen2018participation}. This bias is problematic for two reasons, one of interest to researchers and the other of interest to instructors. For researchers, the analyses above may not be accurate when taking into account a larger population of students (including lower-performing students). Instructors using the PLIC in their classes who achieve high participation rates may be led to incorrectly conclude that their students performed below average because their class included a larger proportion of low-performing students. Using imputation, we quantified the bias in the matched dataset to more accurately represent PLIC scores across a wider group of students and to be transparent to instructors wishing to use the PLIC in their classes.

Imputation is a technique for filling in missing values in a dataset with plausible values for more complete datasets. In the case of the PLIC, we imputed data for respondents who completed the closed-response version of either the pre- or post-survey, but not both (2173 respondents). The number of PLIC closed-response surveys collected is summarized in Table~\ref{PLIC_CR} for both pre- and post-surveys. The 245 students in Table~\ref{PLIC_CR} who are missing both pre- and post-survey data represent the students who completed at least one open-response survey, but no closed-response surveys. Without any information about how these students perform on a closed-response PLIC survey or other useful information such as their grades or scores on standardized assessments, these data cannot be reliably imputed.
\begin{table}[htbp]
\caption{Number of PLIC closed-response pre- and post-surveys collected.\label{PLIC_CR}}
\begin{ruledtabular}
\begin{tabular}{p{1.6cm} c c}
& \textbf{Post-survey} & \textbf{Post-survey}\\
& \textbf{missing} & \textbf{completed}\\
\hline
\multirow{2}{1.6cm}{Pre-survey missing} & \multirow{2}{*}{245} & \multirow{2}{*}{797}\\\\
\multirow{2}{1.6cm}{Pre-survey completed} & \multirow{2}{*}{1376} & \multirow{2}{*}{1911}\\\\
\end{tabular}
\end{ruledtabular}
\end{table}

We used Multivariate Imputations by Chained Equations (MICE)~\cite{MICE} with predictive mean matching (pmm) to impute the missing closed-response data. These methods have been discussed previously in~\citep{nissen2018missing, van2018flexible} and so we do not elaborate on them here. For each respondent, we used the levels of the lab they were enrolled in (FY or BFY), the type of lab they were enrolled in (CTLabs or other), and the score on the closed-response survey they completed to estimate their missing score. MICE operates by imputing our dataset $M$ times, creating $M$ complete datasets, each containing data from 4084 students. Each of these $M$ datasets will have somewhat different values for the imputed data~\citep{nissen2018missing, van2018flexible}. If the calculation is not prohibitive, it has been recommended that $M$ be set to the average percentage of missing data~\citep{van2018flexible}, which in our case is 27. After constructing our $M$ imputed datasets, we conducted analyses (means, \textit{t}-tests, effect sizes, regressions) on these datasets separately, then combined our $M$ results using Rubin's rules to calculate standard errors~\citep{rubin1976inference}.

Using our imputed datasets, we now demonstrate that the results shown above concerning the overall scores and measures of concurrent validity are largely the same as with the imputed data set. We did not examine measures involving individual questions or their correlations (question scores, discrimination, internal consistency) as the variability between questions makes the predictions through imputation unreliable. The test-retest reliability measure included all valid closed-response pre-survey responses, and so there is much less missing data, making the imputation unnecessary.

\subsection{Test scores}
Mean scores for the matched, total, and imputed datasets are shown in Table~\ref{MI_Means}. The pre-survey mean of the imputed dataset is not statistically different from the pre-survey mean of the matched dataset or the valid pre-surveys dataset. Similarly, the post-survey mean of the imputed dataset is not statistically different from the post-survey mean of the valid-post surveys dataset. There is, however, a statistically significant difference in post-survey means between the matched and imputed datasets ($t$-test, $p < 0.05$) with a small effect size (Cohen's $d = 0.08$). Additionally, as with the matched dataset, there is a statistically significant difference between pre- and post-means in the imputed dataset ($t$-test, $p < 0.001$). However, the effect size is smaller (Cohen's $d = 0.16$) than in the matched dataset.

\begin{table}[htbp]
\caption{Average scores for datasets containing only matched students, all valid responses, and the complete set of students including imputed scores.\label{MI_Means}}
\begin{ruledtabular}
\begin{tabular}{cl c c c}
\textbf{Dataset} & \textbf{N} & \textbf{Pre Avg.} & \textbf{Post Avg.}\\ 
\hline
Matched & 1911 & $5.25\pm0.05$ & $5.52\pm0.05$\\
Valid Pre-surveys & 3287 & $5.21\pm0.04$ &  \\
Valid Post-surveys & 2708 & & $5.43\pm0.05$ \\
Imputed & 4046 & $5.20\pm0.04$ & $5.42\pm0.04$ \\
\end{tabular}
\end{ruledtabular}
\end{table}

\subsection{Concurrent Validity}

Our imputed dataset contains 3428 respondents from FY labs and 656 respondents from BFY labs. Because the experts only filled out the survey once, there is no missing data and imputation is unnecessary. In Table~\ref{Score_Level_MI}, we report again the average scores for students enrolled in FY and BFY level physics lab courses and experts.

\begin{table}[htbp]
\caption{Scores across levels of physics maturity, where \textit{N} is the number of responses in the imputed dataset (except for experts). Significance levels and effect sizes are reported for differences in pre- and post-test means. (FY = first-year lab course, BFY = beyond-first-year lab course)\label{Score_Level_MI}}
\begin{ruledtabular}
\begin{tabular}{cl c c c c}
& \textit{\textbf{N}} & \textbf{Pre Avg.} & \textbf{Post Avg.} & \textit{\textbf{p}} & \textit{\textbf{d}}\\ 
\hline
FY & 3428 & $5.12\pm0.04$ & $5.36\pm0.05$ & $<0.001$ & 0.173\\
BFY & 656 & $5.62\pm0.10$ & $5.74\pm0.10$ & 0.052 & 0.088\\
Experts & \multirow{2}{*}{78} & \multirow{2}{*}{$7.6\pm0.2$} & &\\
(non-imputed) & & & &\\
\end{tabular}
\end{ruledtabular}
\end{table}

As in the matched dataset, there is a statistically significant difference in pre- and post-means for students trained in FY labs, but the effect size is smaller. Again, there is no statistically significant difference in pre- and post-means for students trained in BFY labs. Again, an ANOVA comparing pre-survey scores across physics maturity indicates that physics maturity is a statistically significant predictor or pre-survey scores ($F(2, 6210.5) = 218.3$, $p < 0.001)$ with a large effect size ($\eta^{2} = 0.101$). Our imputed dataset contained a total of 505 students who participated in FY CTLabs and 2923 students who participated in other FY physics labs. The linear fit for post-scores using pre-scores and lab type as predictors (see Eq.~\ref{eq:LM}) again found that students trained in CTLabs outperform their counterparts in other labs. Coefficients and significance levels are reported in Table~\ref{tab:LM_MI}.

\begin{table}[htbp]
\caption{Linear model for PLIC post-score as a function of lab treatment and pre-score using imputed dataset.\label{tab:LM_MI}}
\begin{ruledtabular}
\begin{tabular}{l c c c}
\textbf{Variable} & \textbf{Coefficient} & \textit{\textbf{p}}\\ 
\hline
Constant, $\beta_{0}$ & $4.3\pm0.3$ & $<0.001$\\
CTLabs, $\beta_{1}$ & $0.40\pm0.13$ & $<0.001$\\
Pre-score, $\beta_{2}$ & $0.27\pm0.06$ & $<0.001$\\
\end{tabular}
\end{ruledtabular}
\end{table}

\subsection{Summary and Limitations}
In this appendix, we have demonstrated via multiple imputation that PLIC scores may be, on average, slightly lower than those reported in the main article. This is likely due to a skew in representation toward higher-performing students in the matched dataset and is more prevalent on the post-survey than the pre-survey. This bias does not, however, affect the conclusions of the concurrent validity section of the main article. There is a statistically significant difference in pre-scores between students enrolled in FY and BFY labs, as well as experts. Students trained in CTLabs score higher on the post-survey, on average, than students trained in other physics labs after taking pre-survey scores into account.

The main limitation to imputing our data in this way stems from the reliability of the imputed values. As briefly mentioned above, we lack information about students' grades or their scores on other standardized assessments, which have been shown to be useful predictors of student scores on diagnostic assessments like the PLIC~\citep{nissen2018missing}. Without this information, the reliability of the imputed dataset is limited. Estimating missing PLIC scores using the predictor variables above (level and type of lab a student was enrolled in and their score on one closed-response survey) likely provides a better estimate of population distributions than simply ignoring the missing data, but much of the variance in scores is not explained by these variables alone.

\bibliography{PLIC2.bib}

\end{document}